\documentclass[lettersize,journal]{IEEEtran}
\usepackage{amsmath,amsfonts,amssymb,amscd}
\usepackage{mathtools}
\usepackage{algorithmic}
\usepackage{algorithm}
\usepackage{array}
\usepackage{isomath}
\usepackage[caption=false,font=normalsize,labelfont=sf,textfont=sf]{subfig}
\usepackage{threeparttable}
\usepackage{makecell}
\usepackage{textcomp}
\usepackage{stfloats}
\usepackage{url}
\usepackage{verbatim}
\usepackage{graphicx}
\usepackage{cite}
\usepackage{cleveref}
\crefformat{section}{\S#2#1#3}
\crefformat{subsection}{\S#2#1#3}
\crefformat{subsubsection}{\S#2#1#3}
\crefrangeformat{section}{\S\S#3#1#4 to~#5#2#6}
\crefmultiformat{section}{\S\S#2#1#3}{ and~#2#1#3}{, #2#1#3}{ and~#2#1#3}
\usepackage{siunitx}
\DeclareMathAlphabet{\mathsfit}{T1}{\sfdefault}{\mddefault}{\sldefault}
\SetMathAlphabet{\mathsfit}{bold}{T1}{\sfdefault}{\bfdefault}{\sldefault}
\usepackage{isomath}
\usepackage{textgreek}
\usepackage{textcomp}
\usepackage{xcolor}
\usepackage[bold]{hhtensor}
\usepackage{textcomp}
\newcommand{\tb}{\textcolor{black}}

\begin{document}

\title{Characterization of Transmission Lines in Microelectronics Circuits using the ARTEMIS Solver}

\author{Saurabh S. Sawant, Zhi Yao, Revathi Jambunathan, and Andrew Nonaka}

\author{Saurabh S. Sawant,
        Zhi Yao,~\IEEEmembership{Member,~IEEE,}
        Revathi Jambunathan,\,
        and Andrew Nonaka
        
\thanks{Manuscript was created in July, 2022; This material is based upon work supported by the U.S.~Department of Energy, Office of Science, the Microelectronics Co-Design Research Program, under contract no. DE-AC02-05-CH11231.}
\thanks{S. S. Sawant, Z. Yao, R. Jambunathan, A. Nonaka are with the Center for Computational Sciences and Engineering at Lawrence Berkeley National Laboratory, Berkeley, California, USA (email: saurabhsawant@lbl.gov; jackie\_zhiyao@lbl.gov; rjambunathan@lbl.gov; ajnonaka@lbl.gov).}} 


\maketitle

\begin{abstract}
    Modeling and characterization of electromagnetic wave interactions with \tb{microelectronic} devices to derive network parameters has been a widely used practice in the electronic industry.  However, as these devices become increasingly \tb{miniaturized} with finer-scale geometric features, computational tools must make use of manycore/GPU architectures to efficiently resolve length and time scales of interest.  This has been the focus of our open-source solver, ARTEMIS (Adaptive mesh Refinement Time-domain ElectrodynaMIcs Solver), which is performant on modern \tb{GPU-based} supercomputing architectures while being amenable to additional physics coupling. This work demonstrates its use for characterizing network parameters of transmission lines using established techniques.  A rigorous verification and validation of the workflow is carried out, followed by its application for analyzing a \tb{transmission line on a CMOS chip designed for a photon-detector application.} Simulations are performed for millions of timesteps on state-of-the-art GPU resources to resolve nanoscale features at gigahertz frequencies.  The network parameters are used to obtain phase delay and characteristic impedance that serve as inputs to SPICE models.  \tb{The code is demonstrated to exhibit ideal weak scaling efficiency up to 1024 GPUs and 84\% efficiency for 2048 GPUs,} which underscores its use for network analysis of larger, more complex circuit devices in the future.

\end{abstract}


\section{Introduction}
Characterization of transmission lines in terms of network parameters is a widely used practice in the high frequency industry for quantifying the performance of circuit components under the interaction of microwaves~\cite{taflove3rd}.
The scattering matrix or $\mathsfbfit{S}$-parameter \tb{matrix} is typically used as a performance metric that connect full-wave field analysis with circuit designs \cite{Mongia_1st,pozar4th,hall1st}.  
Existing computational tools to obtain $\mathsfbfit{S}$-\tb{parameters} use spectral methods~\cite{HFSS,COMSOL} or the finite difference time domain (FDTD) method~\cite{zhang88_MSValid,sheen90_LPFValid} for solving the time-dependent Maxwell's equations.
In the present work we use the latter to characterize performance of \tb{part of} a microscale \tb{transmission line, shown in Fig.~\ref{f:CodCircuitSketch}, that will connect carbon nanotubes to integrated circuit (IC) inputs on a complementary metal-oxide-semiconductor (CMOS) chip that is being designed for a novel photodetector application~\cite{Francois_2022}. 
In principle, carbon nanotubes will be functionalized with quantum dots that absorb photons of different wavelengths depending on their size and produce pulses that traverse through the transmission line. In this work, we will not model carbon nanotubes or study the process of pulse generation, but focus only on the characterization of scattering matrix of this line.}
The scattering matrix will in turn be used to obtain characteristic impedance and phase delay, which will inform higher-level SPICE circuit simulators\cite{Nagel}.
The electromagnetic pulses flowing through the circuit will have frequencies ranging from megahertz to hundreds of gigahertz; however, in this work, we will limit our attention only to the upper end of the frequency range.

The \tb{conventional} FDTD method typically requires computational cell resolution that is an order of magnitude less than the finest geometrical feature in the device; in the present case we consider a resolution as small as $\sim$2-5~nm, limiting the computational timestep to the order of 10~attoseconds.
With such a small timestep, resolving frequencies down to tens of gigahertz requires millions of timesteps.
Naturally, we need a computational solver that efficiently uses the next-generation multicore/GPU exascale architectures.
Existing commercial applications such as \tb{ANSYS HFSS}\textsuperscript{\textregistered}~\cite{HFSS,HFSSDevelpment} offer limited scalability while requiring license fees.
For example, the HFSS solver has demonstrated only 2x speed up with GPUs compared to CPUs~\cite{HFSS_HPC_Blog}. 
Also, the lack of user access to the source code means these applications are not adaptable for today's beyond-CMOS research needs, which are increasingly geared toward investigations of coupled physics phenomena involving novel materials such as superconductors. 
In this work, we employ our open-source FDTD solver, known as ARTEMIS (Adaptive mesh Refinement Time-domain ElectrodynaMIcs Solver)~\cite{zhi22_1stArtemis}, which has been recently demonstrated to exhibit \tb{59x} speedup on the NVIDIA A100 GPU nodes with near perfect scaling on modern \tb{supercomputing} architectures such as the Perlmutter system at the National Energy Research Scientific Computing Center (NERSC)~\cite{Perlmutter} and shown to be portable across platforms ranging from laptops to manycore/GPU exascale architectures. 
\tb{We note that comparing ARTEMIS to the commercial XFdtd\textsuperscript{\textregistered}~\cite{XFdtd} software, the run times are nearly identical on an AMD Ryzen\textsuperscript{TM} 9 3900X processor, yet, as we will show in this work, the true strength of ARTEMIS is in its capability to run and scale well on modern GPU architectures.}

\tb{Many researchers have proposed modifications to the conventional FDTD method for the applications of modeling large-scale interconnects to circumvent the limitations imposed by Courant-Friedrichs-Lewy (CFL) stability condition, e.g. alternating-direction implicit FDTD~\cite{Lee_and_Chen_ADI_FDTD,Namiki_ADI_FDTD}, split-step FDTD~\cite{wang2013unconditional,kong2013two}, Crank-Nicolson FDTD~\cite{sun2003unconditionally,tan2008efficient}, and others~\cite{jia2008arbitrary,afrooz2011efficient,jia2008arbitrary}. 
Recently, Kumar {\it et al.}~\cite{Kumar_UncondStableFDTD} have proposed an unconditionally stable FDTD method, which they used to model crosstalk effects of the very large scale integration (VLSI) interconnects and through silicon vias (TSVs)~\cite{RAMESHKUMAR2015155,KUMAR20151263,Kumar_GrapheneBasedTSVs}.
In this work, we have not explored the applicability of these advanced FDTD methods for solving Maxwell's equations and have focused on demonstrating the use of GPU-enabled ARTEMIS for $\mathsfbfit{S}$-matrix calculations, setting the framework for future analysis.}

The paper is organized as follows: \cref{sec:Methodology_SandRLCG} describes the methodology employed to compute the scattering matrix. 
In \cref{sec:Microstrip}, the workflow is verified by comparing the scattering matrix, inductance, and capacitance of a simple microstrip circuit obtained from ARTEMIS with the theory and validated by comparing the return and insertion losses for a low pass filter.
\cref{sec:CodesignCircuit} is devoted to computations of the scattering matrix of the \tb{new transmission line} and evaluation of the overall characteristic impedance and phase delay.
\tb{\cref{sec:weakscaling} demonstrates the weak scaling of ARTEMIS on NVIDIA A100 GPUs on the Perlmutter supercomputer~\cite{Perlmutter}.}
A brief summary of the work is provided in \cref{sec:summary} along with comments on future work.

\section{Characterization of Transmission Lines using ARTEMIS}~\label{sec:Workflow}
\subsection{Methodology to compute scattering parameter}~\label{sec:Methodology_SandRLCG}
One of the typical computational approaches to obtain the $\mathsfbfit{S}$-\tb{parameters} includes excitation of time-varying pulses of broadband frequencies at various ports one by one, collection of diagnostics such as temporal histories of voltages and currents at these ports, evaluation of discrete Fourier transforms (DFTs) of these signals, and then taking their ratios in Fourier space to evaluate frequency-dependent parameter components~\cite{taflove3rd}. 
The scattering matrix obtained in this way is for matched loading conditions, denoted here by $\mathsfbfit{S}^*$, which is then transformed to the $\mathsfbfit{S}$ matrix normalized to 50~\textOmega~reference port impedances.
$\mathsfbfit{S}^*$ can also be transformed to other network parameters, which can, in turn, be used to extract the characteristic impedance of the overall circuit and lumped circuit elements. 
These steps are briefly described next.

We follow the procedure of Sheen {\it et al.}~\cite{sheen90_LPFValid} to compute components of the power-mode scattering matrix~\cite{oh_1st} as
\begin{align}
    \mathsfit{S}_{ij}^{*}(f, y_i, y_j) &=
    \begin{dcases}
            \frac{\mathcal{F}[V_{\text{ref}}(y_j,t)]}{\mathcal{F}[V_{\text{inc}}(y_i,t)]}  & \text{if} \quad i = j\\[0.6em]
            \frac{\mathcal{F}[V_{\text{tra}}(y_i,t)]}{\mathcal{F}[V_{\text{inc}}(y_j,t)]}  \sqrt{\frac{Z_{r}(y_j,f)}{Z_{r}(y_i,f)}} & \text{if} \quad i \ne j
    \end{dcases}
    \label{eq:S_Pmode}\\
    \intertext{where}
       \mathcal{F}[V_{\text{ref}}(y_j,t)] &= \mathcal{F}[V_{\text{tot}}(y_j,t)] - \mathcal{F}[V_{\text{inc}}(y_j,t)] \nonumber
\end{align}
Here $\mathcal{F}$ indicates the discrete Fourier transform of \tb{the} voltage pulse varying with time $t$, $Z_{r}$ is the reference impedance of a port calculated as 
\begin{align}
        Z_{r}(y_j,f) &= \frac{\mathcal{F}[V_{\text{inc}}(y_j,t)] }{ \mathcal{F}[I_{\text{inc}}(y_j,t)]}
        \label{eq:Zr}
\end{align}
and $V_{\text{inc}}$, $V_{\text{tot}}$, $V_{\text{ref}}$, and $V_{\text{tra}}$ are the incident, total, reflected, and transmitted voltages, which will be defined shortly.
In this approach, two separate simulations are carried out for each port $j$, designated as a driving port.
In the first simulation, a voltage wave incident at port $j$, referred to by the subscript `inc', is modeled without any internal or boundary reflections.
This is done by making the portion of the circuit in the vicinity of the port uniform in the propagation direction and imposing perfectly matched layer (PML) boundary condition~\cite{berenger94_PML,berenger96_3DPML} on all boundaries other than the ground.
The ground is modeled as a perfect electric conductor (PEC) by imposing transverse electric fields on the boundary to be zero.
In the second simulation, the same pulse is introduced at port $j$ and the entire circuit with all internal discontinuities is modeled while imposing the same boundary conditions as before.
Since the incident pulse at port $j$ may coexist with the reflected pulse, the pulse measured at port $j$ is denoted as the total pulse, referred to by the subscript `tot'.
From the total and incident portions of the pulse, the reflected portion of the pulse is computed and denoted by the subscript `ref'.
The pulses registered at all other ports $i$ are denoted as transmitted pulses, denoted by the subscript `tra'.

A transverse electromagnetic (TEM) wave is excited by introducing a differential Gaussian pulse of the $z$-component of the electric field, $E_z$, given as
\begin{align}
    E_z(y_j,t) &=  -2\frac{(t-t_0)}{\sigma} \exp\left( - \frac{(t-t_0)^2}{\sigma^2}\right)
        \label{eq:EzExcitation}
\end{align}
where $\sigma$ is the standard deviation of the pulse, which is related to the frequency at which the pulse has the highest magnitude, $f_p = (\sqrt{2} \pi \sigma)^{-1}$, and $t_0$ is the time at which the pulse peaks, which is set equal to $3\sigma$ to start the pulse smoothly.
The pulse is applied uniformly on the $xy$ edge-centered points of the Yee-grid~\cite{yee66} located on the $y=0$ plane bounded by two metal conductors, as indicated by surface `e' in Fig.~\ref{f:Microstrip} for the microstrip \tb{line}.
A soft source is used to avoid undesirable retroreflections from internal parts of the circuit from disrupting the incident signal~\cite{taflove3rd}. 

Time-dependent voltage and current, $V$ and $I$, respectively, at location $y_i$ along the propagation direction are obtained as
\begin{equation}
    \begin{split}
        V(y_i,t) &=  \frac{1}{w}\int \vec{E} \cdot d\vec{A} = \frac{\iint_e E_z(y_i,t) dx }{w}\\
            I(y_i,t) &= \oint \vec{H} \cdot d\vec{l}  =  - \int_{a} H_x(y_i,t) dx \\
                     &+ \int_b H_z(y_i,t) dz + \int_c H_x(y_i,t) dx  
                     - \int_d H_z(y_i,t) dz  \\
    \end{split}
    \label{eq:VandI}
\end{equation}
For computing voltage accurately, the surface integral of $E_z$ is evaluated over a surface bounded by the two metals, such as surface $e$, and the integral is divided by the width of the internal conductor, $w$.
Current, $I$, is evaluated by taking four line-integrals of the $\vec{H}$-field in the clockwise direction around the internal metal, as shown in the insert of Fig.~\ref{f:Microstrip} with arrows passing through face-centered Yee-grid cells that are a part of  surfaces `a', `b', `c', and `d' around the metal one cell distance away.

The $\mathsfbfit{S}^*$-matrix is transformed to the $\mathsfbfit{S}$-matrix normalized to reference impedances of 50~\textOmega~using the procedure of Reveyrand~\cite{reveyrand18_paramConv} as
\begin{align}
            \mathsfbfit{S} &= \mathsfbfit{A}^{-1} \cdot (\mathsfbfit{S}^* - \mathsfbfit{B^*} ) \cdot (\mathsfbfit{I} - \mathsfbfit{B} \cdot \mathsfbfit{S}^*)^{-1} \cdot \mathsfbfit{A^*} \label{eq:STransformation} \\
\intertext{where}
            \mathsfbfit{A} &= \mathsfbfit{G^*}^{-1} \cdot \mathsfbfit{G} \cdot (\mathsfbfit{I} - \mathsfbfit{B^*})  \nonumber\\
            \mathsfbfit{B} &= (\mathsfbfit{Z_r^*} - \mathsfbfit{Z_r}) \cdot (\mathsfbfit{Z_r^*} - \mathsfbfit{Z_r})^{-1} \nonumber
\end{align}
In (\ref{eq:STransformation}), all matrices are of order $N$ corresponding to an $N$-port network, $\mathsfbfit{I}$ is an identity matrix, and $\mathsfbfit{G}$, $\mathsfbfit{Z_r}$, $\mathsfbfit{G^*}$, and $\mathsfbfit{Z_r^*}$ are diagonal matrices with elements
\begin{equation}
        \begin{split}
            \begin{rcases}
                    \mathsfit{G}_{ij} &=  \sqrt{|\text{Re}[Z_{r}(y_i)]|}^{-1} \\
                    \mathsfit{Z}_{r,ij} &=  Z_{r}(y_i) \\
                    \mathsfit{G}_{ij}^* &=  \sqrt{50}^{-1} \\
                    \mathsfit{Z}_{r,ij}^* &=  50 \\
            \end{rcases}
            \text{if} \quad i=j \\
        \end{split}
        \label{eq:RefsforTranform}
\end{equation}

The $\mathsfbfit{S}^*$-parameter can also be converted to an ABCD matrix and used to compute the overall characteristic impedance and lumped RLCG circuit elements using the simplest Telegrapher's model, as described in Refs.~\cite{oh_1st,chu2015robust}.

\subsection{Verification and Validation of the Workflow for $\mathsfbfit{S}$-parameter Computation}~\label{sec:Microstrip}
\begin{figure*}[!t]
    \centering
        \subfloat[]{\label{f:Microstrip}{\includegraphics[width=0.9\textwidth]{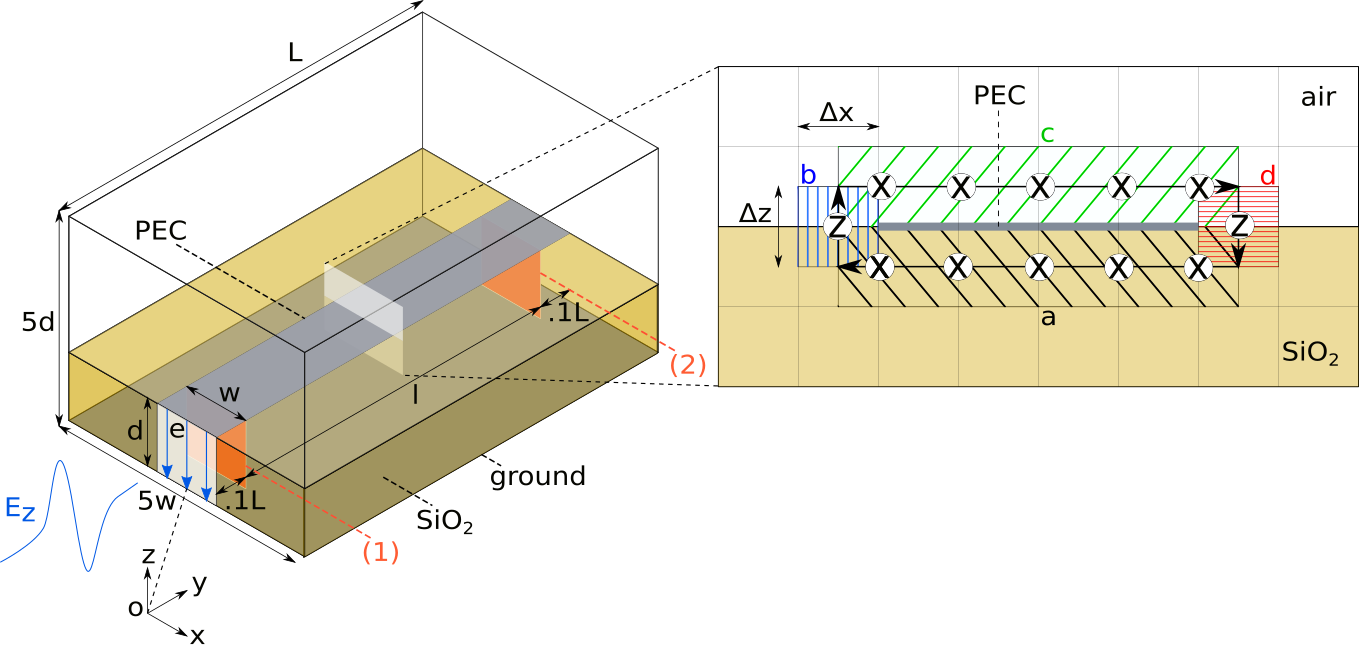}}}\\
        \subfloat[]{\label{f:MS_Sstar}{\includegraphics[width=0.44\textwidth]{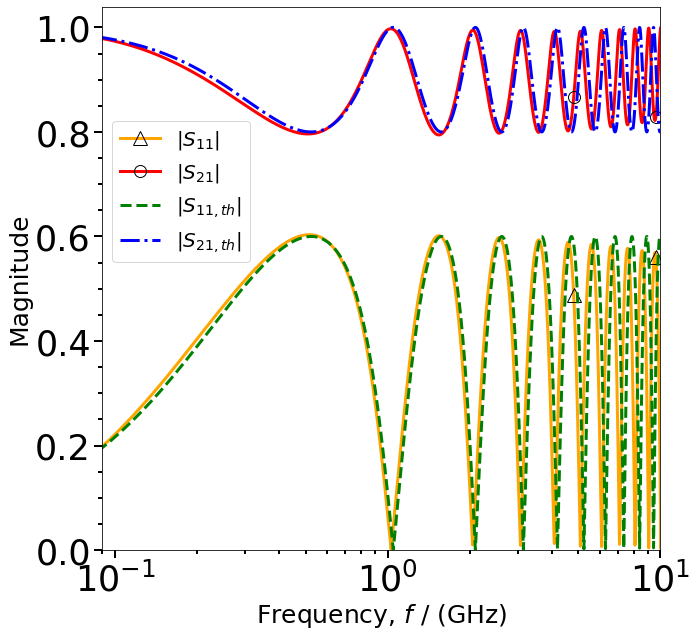}}}\hfill
        \subfloat[]{\label{f:MS_LandC}{\includegraphics[width=0.44\textwidth]{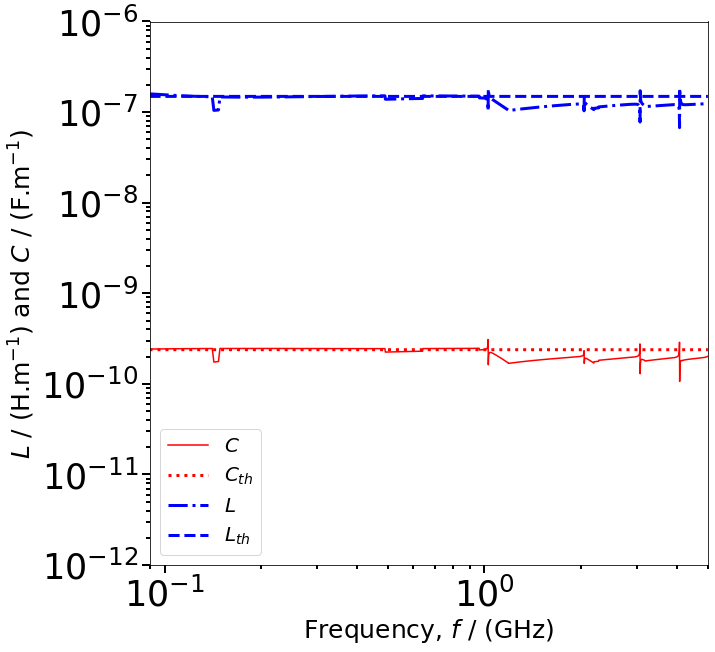}}}\,
    \caption{(\textit{a}) Three dimensional sketch of a microstrip is shown where the metal line is modeled as a PEC. The insert shows surfaces on the Yee-grid around the metal one cell distance away that are used to compute current. `O' indicates the origin whose exact location is indicated by a dashed line, port locations are labeled as (1) and (2), and the excitation plane `e' is overlaid with electric field lines. (\textit{b}) and (\textit{c}) show comparisons of the components of $\mathsfbfit{S}$-matrix and per unit length $C$ and $L$ for a microstrip with the theory.}
\end{figure*}
First, we verify the work flow described in \cref{sec:Methodology_SandRLCG} by applying it to the simplest two-port microstrip circuit, shown in Fig.~\ref{f:Microstrip}.
The microstrip is designed such that its characteristic impedance is $Z_{0,{\rm th}}$=25~\textOmega, the relative dielectric permittivity is $\epsilon_r=3.8$ for silicon dioxide (SiO$_2$), the ratio of the strip width to dielectric thickness is 5.7, dielectric thickness is 1~mm, and the length of the microstrip is 100 times the dielectric thickness.
The simulation domain is discretized on a Yee-grid composed of 200, 1000, and 50 uniform cells in the $x$, $y$, and $z$ directions, respectively.
The Courant–Friedrichs–Lewy (CFL) number of 0.9 is used, with corresponding timestep of 0.173~ps.
The differential Gaussian pulse used for excitation has a pulse width of $\sigma = 11.254$~ps and $t_0 = 3 \sigma$, such that the highest magnitude of the pulse is at a frequency of $f_p=20$~GHz.
The simulation was carried out for 60,000 timesteps, so that the lowest resolved frequency is approximately 0.1~GHz.

Figs.~\ref{f:MS_Sstar} and \ref{f:MS_LandC} show excellent comparisons of components of $\mathsfbfit{S}$ matrix and per unit length inductance and capacitance obtained from ARTEMIS with corresponding theoretical estimates for a lossless transmission line~\cite{SParamTheory,pozar4th} given as
\begin{equation}
    \begin{split}
            \mathsfit{S}_{11,\rm th} &= \frac{(1-\kappa^2)\Gamma}{1-\kappa^2 \Gamma^2}\\
            \mathsfit{S}_{21,\rm th} &= \frac{(1-\kappa^2) \kappa}{(1-\kappa^2 \Gamma^2)} \\
             C_{\rm th} &= \frac{1}{Z_{0,{\rm th}} v_{\rm th}} \\
             L_{\rm th} &= \frac{Z_{0,{\rm th}}}{v_{\rm th}}
    \end{split}
    \label{eq:MS_S_th}
\end{equation}
where $\kappa=\exp(j \beta_{\rm th} l)$, $j$ is the imaginary unit, $\beta_{\rm th}$ is the theoretical propagation constant, $l$ is the distance between two ports, and $\Gamma=-0.333$ is the reflection coefficient at port 2 due to mismatch between 25 and 50~\textOmega~impedances, $v_{\rm th}$ is the theoretical phase velocity, $c$ is the speed of light, and $\epsilon_e$ is the effective dielectric constant.

\begin{figure*}[!t]
    \centering
        \subfloat[]{\label{f:SheenLPF}{\includegraphics[width=0.4\textwidth]{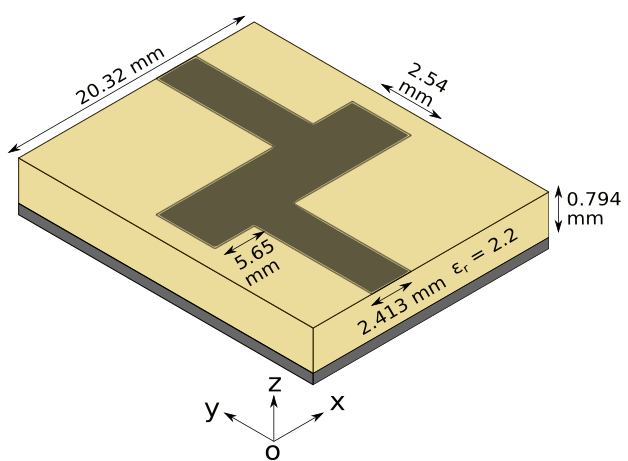}}}\\
        \subfloat[]{\label{f:SheenS11}{\includegraphics[width=0.49\textwidth]{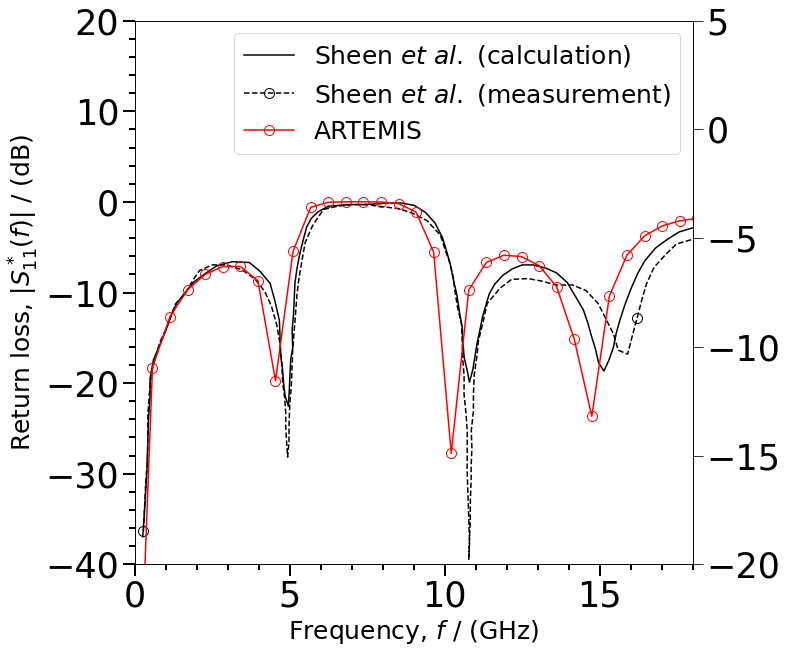}}}\hfill
        \subfloat[]{\label{f:SheenS21}{\includegraphics[width=0.44\textwidth]{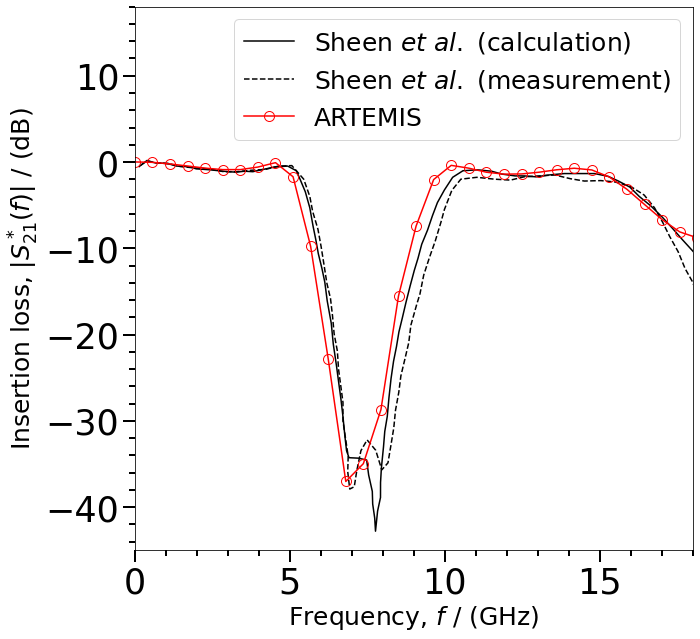}}}\,
    \caption{Validation of scattering matrix computations from ARTEMIS. (\textit{a}) shows the low pass filter of Sheen {\it et al.}~\cite{sheen90_LPFValid} (Fig. 7). (\textit{b}) and (\textit{c}) show the comparisons of the return and insertion losses of the low pass filter with results of Sheen {\it et al.}~\cite{sheen90_LPFValid}.}
\end{figure*}

Next, the $\mathsfbfit{S}$-parameter computation using ARTEMIS is validated for a two-port low-pass filter studied computationally and experimentally by Sheen {\it et al.}~\cite{sheen90_LPFValid}.
The details concerning the structure geometry and the simulation setup can be found in Ref.~\citen{sheen90_LPFValid}.
The good agreements for the return and insertion losses obtained from ARTEMIS are shown in Figs.~\ref{f:SheenS11} and ~\ref{f:SheenS21}, respectively.

\section{\tb{Characteristics of the Newly Proposed Transmission Line}}~\label{sec:CodesignCircuit}
\begin{figure}[!t]
    \centering
        \includegraphics[width=0.44\textwidth]{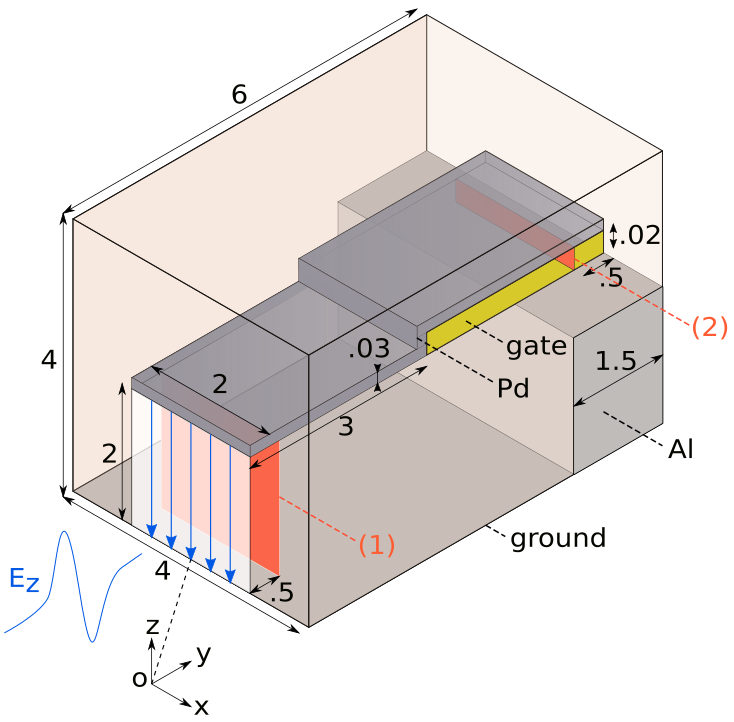}
        \caption{Three dimensional sketch of the \tb{newly proposed transmission line}. Indicated dimensions are in microns and the \tb{oxide} material is assumed to be silicon dioxide (SiO$_2$).}
        \label{f:CodCircuitSketch}
\end{figure}

We now turn to the analysis of the \tb{transmission line} shown in Fig.~\ref{f:CodCircuitSketch}.
The simulation domain is 4, 6, and 4~$\mu$m in the $x$, $y$, and $z$ directions, and consists of a 2~$\mu$m wide and 30~nm thick palladium (Pd) metal stripline at its center spanning the entire length in $y$, while the strip contains a small vertical step discontinuity of 20~nm halfway along the length to make way for the oxide.
In this work, we assume the oxide to be composed of SiO$_2$, although in the final design, it may change.
The line supports a TEM mode which impinges on a \tb{4~$\mu$m} block of aluminum (Al) \tb{located} 4.5~$\mu$m down the line.
The rest of the \tb{line} is filled with SiO$_2$ (not shown in the figure).

The \tb{transmission line} is defined on a Yee-grid composed of 200, 120, and 800 uniform cells in the $x$, $y$, and $z$ directions ($\Delta x=20$~nm, $\Delta y=50$~nm, $\Delta z=5$~nm).
The resolution is highest in the $z$-direction to accommodate four cells in the 20~nm oxide region between the Pd and Al.
Appendix~\ref{app:ConvTests} shows a convergence test that justifies this choice of the grid resolution.
The CFL number of 0.9 is used, which imposes a timestep of 0.0145~fs.
The differential Gaussian pulse has a pulse width of $\sigma = 10.146$~fs and $t_0 = 3 \sigma$, such that it exhibits the highest magnitude at a frequency of $f_p \approx 22.2$~THz, while the magnitude drops by three orders of magnitude to frequencies of 15~GHz and 95~THz.
In total, four simulations are executed to completely obtain the $2 \times 2$ scattering matrix using (\ref{eq:S_Pmode}): two each corresponding to the excitations applied on the $y=0$ and $y=L_y$ boundaries as described in \cref{sec:Methodology_SandRLCG}.
Each simulation is run for 1.4~million time steps such that the lowest resolved frequency for the scattering matrix is 50~GHz.

\begin{figure*}[!t]
    \centering
        \subfloat[]{\label{f:Co_S}{\includegraphics[width=0.44\textwidth]{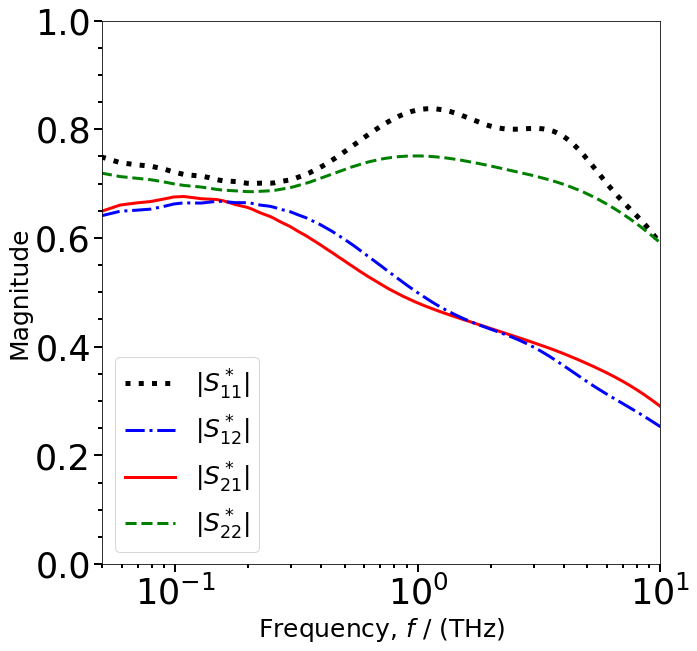}}}\hfill
        \subfloat[]{\label{f:Co_Sstar}{\includegraphics[width=0.44\textwidth]{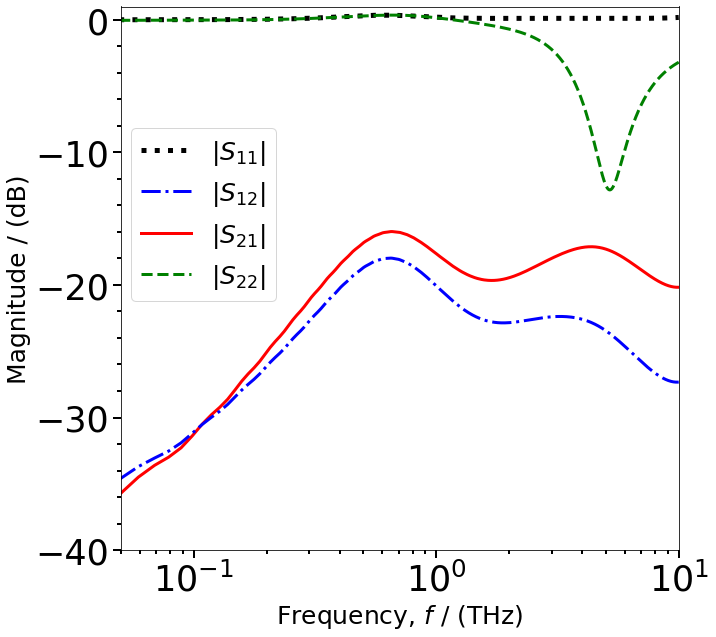}}}\,
        \subfloat[]{\label{f:Co_Z0}{\includegraphics[width=0.44\textwidth]{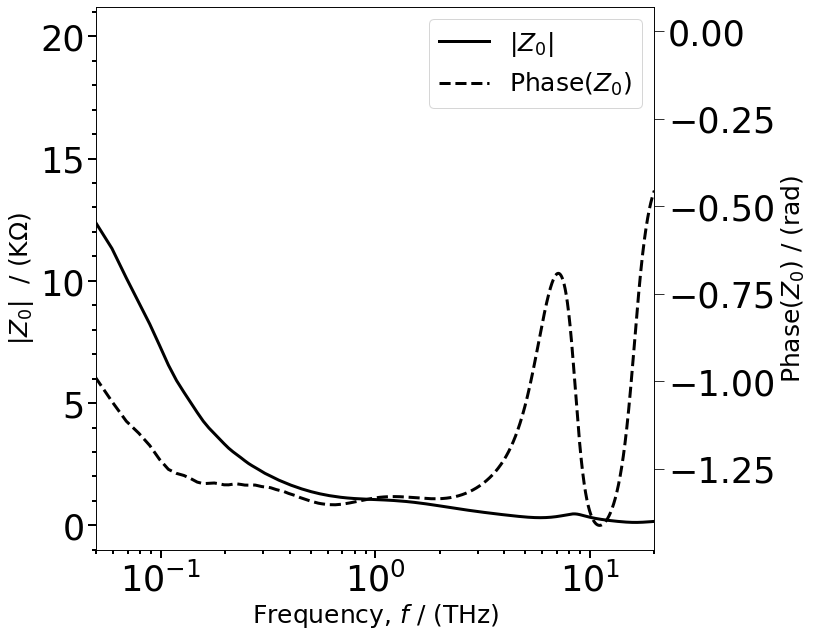}}}\hfill
        \subfloat[]{\label{f:Co_Gamma}{\includegraphics[width=0.44\textwidth]{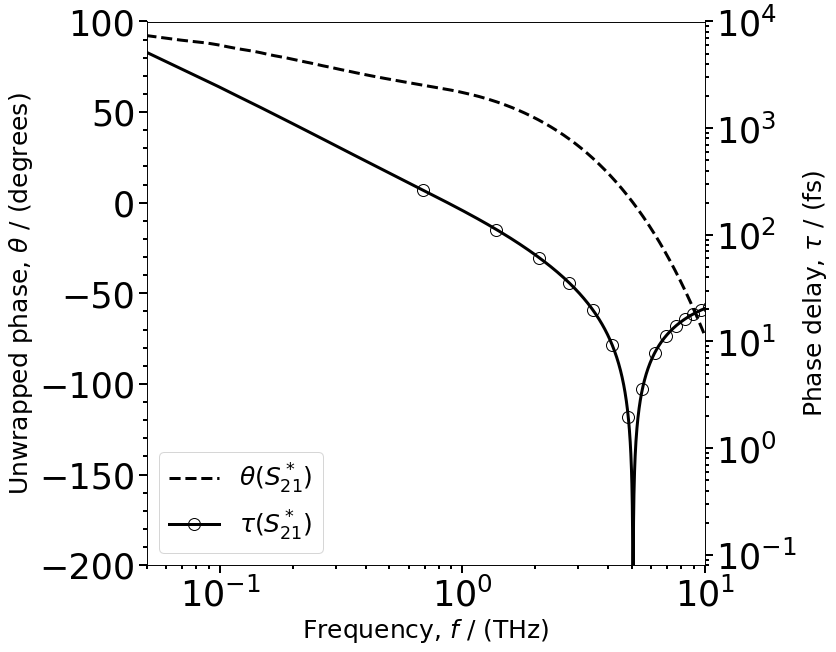}}}\hfill
        \caption{For the \tb{transmission line}, the components of $\mathsfbfit{S}^*$ and $\mathsfbfit{S}$ matrices are shown in (\textit{a}) and (\textit{b}), respectively. The magnitude and phase of the overall characteristic impedance computed from the $\mathsfbfit{ABCD}$-matrix are shown in (\textit{c}), and the unwrapped phase and phase delay of $\mathsfit{S}_{21}^*$ are shown in (\textit{d}).}
\end{figure*}

Fig.~\ref{f:Co_S} shows the $\mathsfit{S}^*$-matrix obtained from (\ref{eq:S_Pmode}), which describes the performance of the circuit under matched load conditions.
This circuit exhibits losses due to metal, dielectric and sharp corners, which become significant as the frequency increases.
For example, at 50~GHz, the fractions of power reflected and transmitted are $|\mathsfit{S}_{11}|^2 = 54.5\%$ and $|\mathsfit{S}_{21}|^2=43.6\%$, respectively, while the rest of $\sim$$1.9\%$ of the power is attributed to losses.
On the other hand, at 10~THz, as high as 56.9\% of the power is lost, while the amount reflected and transmitted are 34.7\% and 8.4\%, respectively.
Next, Fig.~\ref{f:Co_Sstar} shows the $\mathsfbfit{S}$-matrix obtained from $\mathsfbfit{S}^*$ by renormalizing it to 50~\textOmega~reference impedances.
In this case, almost 99.99\% of the power is reflected at all frequencies when the excitation is applied at $y=0$.
This significant reflection is caused primarily by the mismatch between the 50~\textOmega~reference impedance and the high local impedance of the circuit in the vicinity of the excitation plane ($O$(k\textOmega)).

The output obtained from the transmission line analysis serves as \tb{the} input to SPICE circuit models~\cite{hall1st}, either directly or after some simplification depending on the complexity of the model.
For example, the FBLOCK SPICE model directly uses the $\mathsfbfit{S}^*$-parameter, shown in Fig.~\ref{f:Co_S}, at specific discrete frequencies~\cite{Eldo}.
In contrast, some simpler SPICE models use the overall characteristic impedance of the circuit assuming Telegrapher's model and the phase delay in transmission from one port to another.
Here, we provide both of these parameters.
Note that Telegrapher's model can be used only as a crude approximation for the overall characteristic impedance because our \tb{transmission line} is aperiodic in the propagation direction.
Fig.~\ref{f:Co_Z0} shows the magnitude and phase of the overall characteristic impedance, where the former increases rapidly for frequencies below 1~THz.
At 50~GHz, the characteristic impedance is almost 12~k\textOmega.
Next, the phase delay is obtained as~\cite{hall1st} 
\begin{equation}
    \begin{split}
            \tau(\mathsfit{S}_{ij}^*)  = \frac{\theta(\mathsfit{S}_{ij}^*)}{360 f}   \quad  \text{when}  \quad  i \ne j 
    \end{split}
\end{equation}
where $\theta$ is the phase of the scattering component $\mathsfit{S}_{ij}^*$ unwrapped with a period of 2$\pi$.
Fig.~\ref{f:Co_Gamma} shows the unwrapped phase and the phase delay corresponding to $\mathsfit{S}_{21}^*$, while the corresponding quantities for $\mathsfit{S}_{12}^*$ are the same (not shown).
It is seen that at frequencies below $\sim 2$~THz, the phase delay is linear, which implies that in this range of frequencies, there will be no dispersion of the signal during the transmission. 

We close this section with a brief discussion of the computational efforts required to carry out the simulations discussed above.
We used GPU resources available on NERSC's Perlmutter~\cite{Perlmutter}, where one node consists of four NVIDIA A100 GPUs.
Each simulation of 1.4 million time steps on one node ({\it i.e.}, 4 GPUs) takes 41 hours.
By increasing the number of nodes by four times ({\it i.e.}, 16 GPUs) we found that the computational time decreased by a factor of 3.5, {\it i.e.}, the strong scaling speed up is 3.5 (87.5~\% efficiency).
Note that each case was run by decomposing the domain into a number of grids equal to the number of GPUs to optimize the load balancing. 
While strong scaling suggests that our simulations will run faster, to highlight the performance of the code for more complicated transmission line simulations that we plan to perform in the future, we perform weak scaling studies discussed in the next section.

\section{\tb{Weak Scaling Performance Study of ARTEMIS}}~\label{sec:weakscaling}
\begin{figure}[!h]
    \centering
        \includegraphics[width=0.5\textwidth]{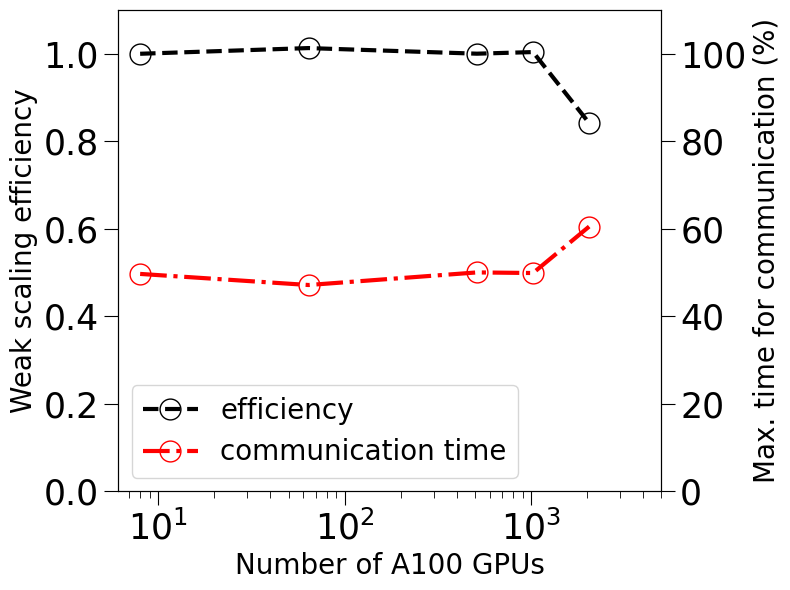}
        \caption{Weak scaling of ARTEMIS using the NVIDIA A100 GPUs on Perlmutter supercomputer.}
        \label{f:weakscaling}
\end{figure}

\tb{In this section, we present the weak scaling study of the FDTD algorithm in ARTEMIS using the NVIDIA A100 GPUs on NERSC's Perlmutter supercomputer.}
\tb{In the weak scaling study, the number of GPUs used for a simulation are increased in equal proportion to the increase in overall computational load, {\it i.e} the total number of computational cells, such that the computational load per GPU remains the same. We quantify the weak scaling efficiency (WSF) as
\begin{equation}
E(N) = \frac{t(8)}{t(N)}
\end{equation}
where $E(N)$ is the WSF for $N$ GPUs, and $t(8)$ and $t(N)$ are the time per timestep averaged over 100 timesteps when 8 and N GPUs are used, respectively. 
Table~\ref{tab:input} lists the number of GPUs and the corresponding computational cells used in five different cases, where the largest case of 2048 GPUs, which amounts to  $33\%$ of the total number of GPUs on Perlmutter, simulated 442.368 billion cells. 
These number of cells per GPU (in this case, $600^3$) are selected such that nearly 80\% of the global memory of each GPU is utilized.
}
under
\tb{Figure~\ref{f:weakscaling} shows that the solver exhibits perfect scaling up to 1024 GPUs, noting that the code spends $\sim$50\% of the time communicating information to fill ghost or guard cells of neighboring grids.
For 2048 GPUs, the efficiency drops to 84.2\%, which is caused by an abrupt 11\% increase in the total communication time. 
Note that in each case, each GPU communicates $\sim$150 MB of total floating point data per time step to neighboring GPUs, and with increase in the number of GPUs the total number of messages to be communicated increase.
We attribute the sudden increase in communication time to the saturation of the system network, which could be reduced in the future by optimizing the way the simulation is run on Perlmutter with the help of system experts at NERSC.}

\begin{table}[H]
\centering
\normalsize
\begin{threeparttable}
\caption{Computational parameters for the weak scaling study.$^a$}
\label{tab:input}       
\begin{tabular}{cccc}
\hline\noalign{\smallskip}
\textbf{Number of GPUs} & \textbf{\makecell{Number of cells in the domain \\(X $\times$ Y $\times$ Z)}}           \\\hline
    8       & 1200 $\times$ 1200 $\times$ 1200                          \\
    64      & 2400 $\times$ 2400 $\times$ 2400                               \\
    512     & 4800 $\times$ 4800 $\times$ 4800                               \\
    1024    & 4800 $\times$ 4800 $\times$ 9600                               \\
    2048    & 4800 $\times$ 9600 $\times$ 9600                              \\
\noalign{\smallskip}\hline\noalign{\smallskip}
\end{tabular}
\begin{tablenotes}
\item $^a$ Number of cells per GPU were 600 $\times$ 600 $\times$ 600 across all runs.
\end{tablenotes}
\end{threeparttable}
\end{table}

\section{Summary and Future Work}~\label{sec:summary}
A general workflow is established using the opensource ARTEMIS solver to obtain the scattering matrix of circuits predominantly supporting a TEM mode.
The workflow is demonstrated by reproducing the theoretical solutions for a simple microstrip and solutions provided by Sheen {\it et al.}~\cite{sheen90_LPFValid} for a low-pass filter.
It is then applied to the \tb{newly proposed transmission line} to extract the power-mode scattering matrix, the characteristic impedance of the overall circuit and the pulse delay that may be used as inputs to the SPICE models. \tb{Finally, the weak scaling study of ARTEMIS is performed using NVIDIA A100 GPUs on NERSC's Perlmutter supercomputer~\cite{Perlmutter}, which revealed 100\% efficiency up to 1024 GPUs and 84\% efficiency for 2048 GPUs, underscoring its capability for performing computationally intensive simulations of intricate transmission lines.}

In the future, we will explore more accurate ways to compute overall characteristic impedance of the circuit from the scattering matrix.
\tb{
We will also extend the applicability of our solver to quantify the crosstalk between multiple transmission line structures.
To investigate $O$(100~nm) long carbon nanotubes and their interaction with quantum dots used in the photon detector CMOS chip, we are also developing a solver for time-dependent nonequilibrium Green's function method, potentially coupled with electrostatic solvers.}

\tb{Other researchers have used the FDTD method to obtain voltages and currents by solving coupled Telegrapher's equations for millimeter-scale multiple transmission lines in VLSI interconnects~\cite{kaushik2006crosstalk,kaushik2007effect,KUMAR2014441,Kumar_UncondStableFDTD}, multi-wall carbon nanotube interconnects~\cite{RAMESHKUMAR2015155,KUMAR20151263,Kaushik_CNTBased3DInterconnects}, and through silicon vias~\cite{Kumar_GrapheneBasedTSVs}.
The underlying FDTD framework of ARTEMIS can be extended for such applications.}
\tb{In these large scale interconnects, process variations play a major role as they can affect the crosstalk and delay~\cite{verma2009effects} and the models to study these can be explored once such framework is built.}

\section*{Acknowledgements}
This work was supported by the US Department of Energy, Office of Science, Office of Basic Energy Sciences, the Microelectronics Co-Design Research Program, under contract no. DE-AC02-05-CH11231 (Co-Design and Integration of nano-sensors on CMOS) for the development of design tools for novel microelectronics.
This research used resources of the National Energy Research Scientific Computing Center (NERSC), a DOE Office of Science User Facility supported by the Office of Science of the U.S. Department of Energy under Contract No. DE-AC02-05CH11231. 
This research leveraged the open-source AMReX code, https://github.com/AMReX-Codes/amrex. 
We acknowledge all AMReX contributors.
Authors thank Dr.~Aikaterini Papadopoulou and Dr.~Maurice Garcia-Sciveres from the Engineering and Physics Divisions, respectively, at the Lawrence Berkeley National Laboratory, and Dr.~Fran\c{c}ois L\'eonard from the Nanoelectronics and Nanophotonics group at the Sandia National Laboratories for valuable discussions.

\section{Appendix: Convergence Test}~\label{app:ConvTests}
\begin{figure}[!t]
    \centering
        \includegraphics[width=0.45\textwidth]{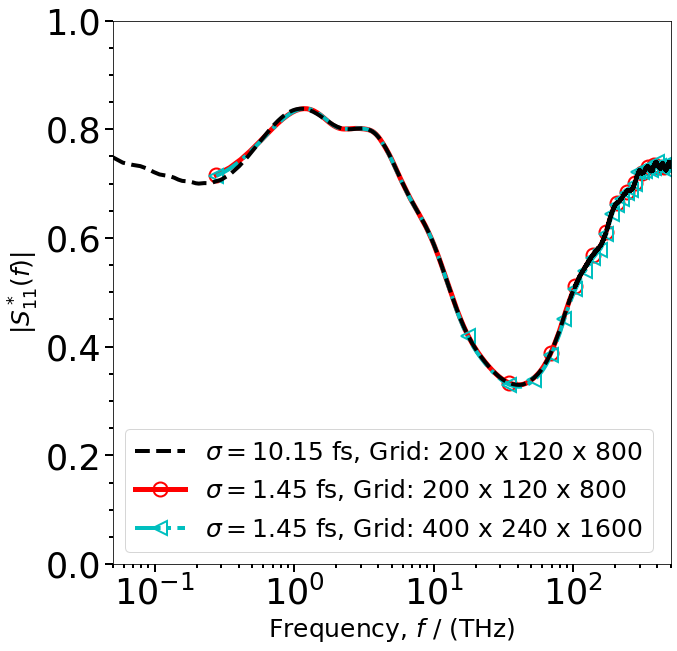}
        \caption{Convergence is demonstrated by comparing the solution of $|\mathsfit{S}_{11}^*|$ obtained on the original grid resolution using the wide pulse with that on the original and finer grids using the narrow pulse.}
        \label{f:GridConv}
\end{figure}
The convergence of the FDTD solution with respect to the chosen grid resolution in \cref{sec:CodesignCircuit} was tested by comparing the magnitude of $\mathsfit{S}_{11}^*$, first shown in Fig.~\ref{f:GridConv}.
However, to reduce the overall computational cost, a shorter simulation was run by modeling a much narrower pulse having the pulse width of $\sigma=1.45$~fs for an overall duration of only 60~k timesteps.
Note that the timestep did not change by the change in the pulse width because all other parameters, as well as the grid resolution, were unchanged, {\it i.e.}~200, 120, 800 Yee-cells in $x$, $y$, and $z$-directions.
This narrow pulse exhibits its highest magnitude at a frequency of $\sim$155.3~THz, while the magnitude drops to three orders at frequencies of 100~GHz and 700~THz.
The total time duration imposes a limit on the lowest accurately resolved frequency, which is 200~GHz for the present simulation.
Fig.~\ref{f:GridConv} shows that for frequencies within 200~GHz to 95~THz, the solutions of $\mathsfit{S}_{11}^*$ computed for both pulses are indistinguishable from each other.
Next, the simulation for the narrow pulse was repeated with double the number of computational cells in each direction, {\it i.e.}~400, 240, 1600 cells in $x$, $y$, and $z$-directions.
Fig.~\ref{f:GridConv} shows that the solution of $\mathsfit{S}_{11}^*$ obtained on the finer grid agrees well with that on the original grid.

\bibliographystyle{IEEEtran}
\bibliography{References}

\newpage

\vfill

\end{document}